\documentclass[% 
reprint,
 aps,reprint,nofootinbib,nobibnotes,notitlepage,superscriptaddress,prl,
 amsmath,amssymb
]{revtex4-2}

\usepackage[caption=false]{subfig}
\usepackage{braket}
\usepackage{float}
\usepackage{lipsum}
\usepackage{graphicx}% include figure files
\usepackage{dcolumn}% align table columns on decimal point
\usepackage{bm}% bold math
\usepackage{natbib}
\usepackage{hyperref}
\hypersetup{
	colorlinks = true,
    linkcolor = Red,
    urlcolor  = Red,
    citecolor = Red
}

\usepackage{microtype}

\usepackage{verbatim}
\usepackage[amssymb]{SIunits}
\usepackage{tabularx}
\usepackage[dvipsnames]{xcolor}
\usepackage{wasysym}

\usepackage{tikz}

\def\be{\begin{equation}}
\def\ee{\end{equation}}
\def\beq{\begin{eqnarray}}
\def\eeq{\end{eqnarray}}

\usepackage{empheq}

\def\mnras{\rm{MNRAS}}

\def\aap{\rm{A\&A}}

\def\apj{\rm{ApJ}}                 % Astrophysical Journal
\usepackage{color}
\definecolor{darkgreen}{RGB}{0,120,0}

\definecolor{darkgreen}{RGB}{0,120,0}

\newcommand{\btheta}{\boldsymbol{\theta}}
\newcommand{\bx}{\bm{x}}
\newcommand{\kmax}{k_{\mathrm{max}}}
\newcommand{\ihmpc}{h^{-1}\mathrm{Mpc}}

\begin{document}

%\preprint{APS/123-QED}

% \title{Hybrid SBI: Combining Perturbation Theory with Simulation Based Inference}
\title{Hybrid SBI or How I Learned to Stop Worrying and Learn the Likelihood}
% 

% \author{A Good Man}
%   \email{emailid@institution.edu}
% \affiliation{%
%  A good place
% }%

% \author{Also a man}
%  \affiliation{
% And a place
% }%

\author{Chirag Modi}
\email{cmodi@flatironinstitute.org}
\affiliation{Center for Computational Astrophysics, Flatiron Institute, New York, NY 10010, USA}
\affiliation{Center for Computational Mathematics, Flatiron Institute, New York, NY 10010, USA}
\author{Oliver~H.\,E.~Philcox}
\email{ohep2@cantab.ac.uk}
\affiliation{Simons Society of Fellows, Simons Foundation, New York, NY 10010, USA}
\affiliation{Center for Theoretical Physics, Columbia University, New York, NY 10027, USA}


\begin{abstract}
\noindent We propose a new framework for the analysis of current and future cosmological surveys, which combines perturbative methods (PT) on large scales with conditional simulation-based implicit inference (SBI) on small scales. This enables modeling of a wide range of statistics across all scales using only small-volume simulations, drastically reducing computational costs, and avoids the assumption of an explicit small-scale likelihood. As a proof-of-principle for this hybrid simulation-based inference (HySBI) approach, we apply it to dark matter density fields and constrain cosmological parameters using both the power spectrum and wavelet coefficients, finding promising results that significantly outperform classical PT methods. We additionally lay out a roadmap for the next steps necessary to implement HySBI on actual survey data, including consideration of bias, systematics, and customized simulations. Our approach provides a realistic way to scale SBI to future survey volumes, avoiding prohibitive computational costs. 

\end{abstract}

\maketitle

%\tableofcontents

%\subsection{Motivation}

%some generic blurb about the next generation of surveys
\noindent Current and upcoming cosmological surveys such as 
the Dark Energy Spectroscopic Instrument~\citep[DESI;][]{desicollaboration2016, desicollaboration2016a, abareshi2022}, 
the ESA {\em Euclid} satellite mission~\citep{laureijs2011},
Vera C. Rubin Observatory (LSST)~\citep{lsst},
and the NASA Nancy Grace Roman Space Telescope~\citep[Roman;][]{spergel2015, wang2022a}, will probe the Universe across a vast array of scales providing high-resolution measurements with the potential to constrain cosmological parameters to unprecedented precision. The past iteration of cosmological data has been analyzed primarily through traditional analytical methods such as perturbation theory (PT); whilst this has been highly successful \citep{Philcox:2021kcw,DAmico:2022osl,Chen:2021wdi}, the approach is necessarily limited to low-order clustering statistics (such as the two- and three-point functions), and to linear and quasi-linear scales. If we wish to fully exploit the information contained within upcoming large scale structure surveys, new methods are needed. 

Recently, simulation-based inference (SBI) has emerged as a promising alternative to overcome the limitations of traditional methods \citep{alsing2018, alsing2019, jeffrey2021, simbigletter, cranmer2020}.
SBI uses numerical simulations to model observables and their summary statistics, before these are combined with neural density estimation methods to efficiently infer cosmological parameters.
As a result, it can be applied to any summary statistic that can be simulated and evaluated, including those deep in the non-linear regime, inaccessible to PT, and those whose likelihoods are far from Gaussian. A major limitation of SBI is that its training requires a large number of high-fidelity simulations, spanning a wide range of cosmologies. This makes it a computationally expensive exercise; a drawback that is further exacerbated by the increasing simulation volumes and resolutions needed to model upcoming data. The largest simulation suite currently available for training SBI for galaxy clustering (\textsc{quijote} simulations \cite{villaescusa-navarro2020}) has moderately low mass resolution and is only $1\,h^{-3}\mathrm{Gpc}^3$ in volume, which is smaller than than the SDSS III-BOSS survey~\citep{dawson2013}. Creating analogous simulations appropriate for the next generation of cosmological surveys is thus likely to be \emph{computationally prohibitive}. 

The primary challenge in scaling up simulations stems from the simultaneous requirements of large volumes, high resolution, and accurate small-scale dynamics. On the largest scales, however, physics is well understood analytically, thus simulation-based approaches are not strictly necessary. In this regime, the density field is close to Gaussian and with quasi-linear perturbations that can be well modeled using perturbation theory. 
Furthermore, traditional statistics like the power spectrum and bispectrum are close to optimal on such scales \cite{Cabass:2023nyo}, thus any more nuanced analysis schemes will not yield significant additional information. It is only the small, non-linear scales where higher-order statistics dominate and thus SBI is needed. The conclusion is clear: if we can combine perturbative methods for analyzing large scales with simulation-based methods for analyzing small scales, we can mitigate the computational cost, and thus make feasible the full analysis of upcoming generations of surveys. 

\begin{figure*}
    \centering
    \includegraphics[width=0.9\textwidth]{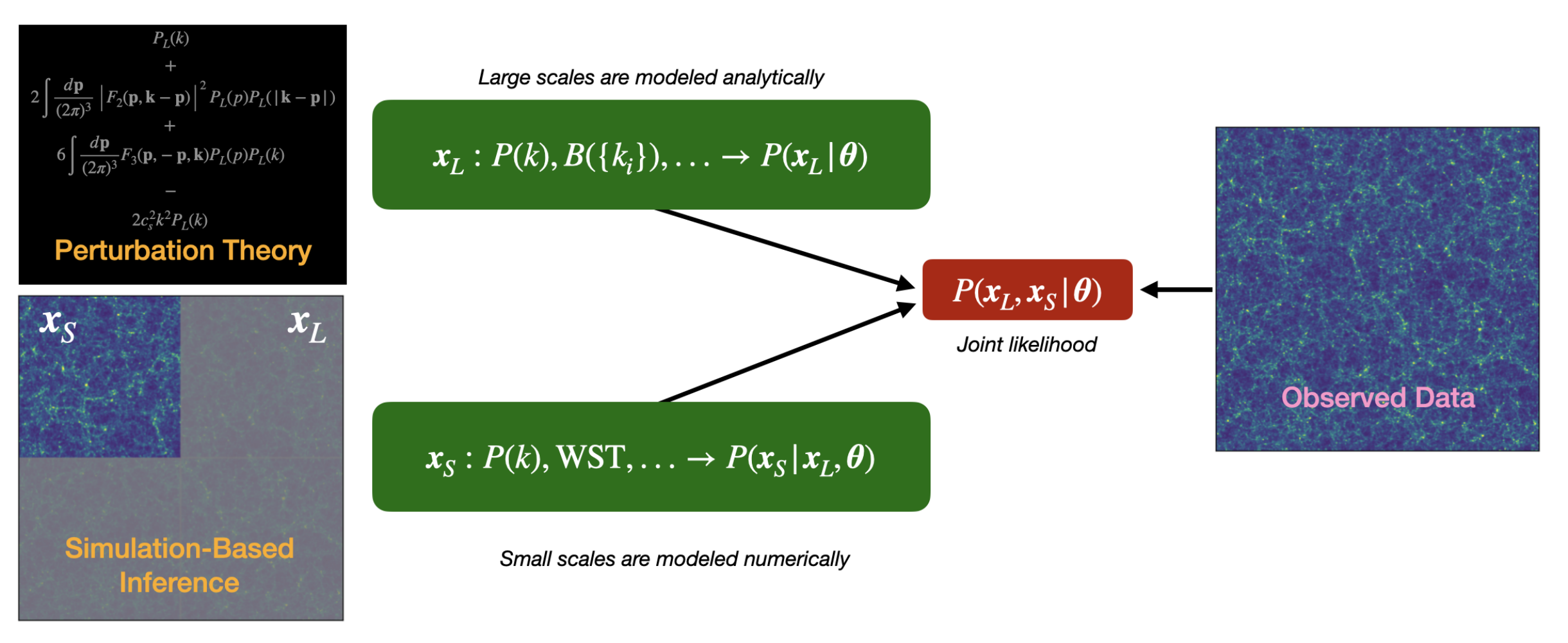}
    \caption{Schematic overview of this work. Large-scale statistics are modeled using classical perturbative techniques, whilst small-scale summaries make use of simulation-based inference, trained on small sub-volume simulations, obviating the need to run costly high-resolution simulations of the entire survey volume.}
    \label{fig: cartoon}
\end{figure*}

In this work, we propose such a framework for hybrid simulation-based inference (HySBI), and demonstrate its utility in predicting cosmological parameters from summary statistics of the matter density field. This is depicted schematically in Fig.\,\ref{fig: cartoon}.
% We begin by reviewing SBI in \section{sec:sbi}

%%%%%%%%%%%%%%%%%%%%%%%%%%%%%%%%%%%%
\section{Simulation-based Inference}
\label{sec:sbi}

\noindent SBI begins with a training dataset composed of ($\btheta, \bx$) pairs where $\btheta$ are the underlying model parameters to be inferred (here, cosmology parameters and any relevant nuisance parameters), and $\bx$ is the corresponding data-vector generated from a simulator (such as an $N$-body realization). Analysis usually proceeds by training a conditional neural density estimator $q_{\phi}$ (with parameters $\phi$) to learn: (a) the likelihood model, $p(\bx|\btheta)$, by maximizing the conditional log-probability of the data $\bx$ conditioned on the parameters $\btheta$ (known as a neural likelihood estimator; NLE); or (b) the posterior itself, $p(\btheta|\bx)$, by maximizing the conditional log-probability of the model parameters $\btheta$ conditioned on the data $\bx$, over the training data (known as a neural posterior estimator; NPE). 

When performing inference on a test observation $\bx'$ when using NPE, we can directly query the trained estimator $q_{\phi^*}$ to generate samples from the posterior, \textit{i.e}\ $\btheta \sim q_{\phi^*}(\btheta|\bx')$. 
In NLE, we instead need to combine the learnt likelihood model $q_{\phi^*}(\bx|\btheta)$ with the prior $p(\btheta)$, then use Markov chain Monte Carlo (MCMC) algorithms to generate samples from this posterior. NPE is often preferred over NLE for two reasons: (1) often the dimensionality of the posterior distribution is smaller than that of the likelihood which can lead to asymptotically better learning; (2) NPE does not require the additional step of running MCMC. In the HySBI framework however, we use NLE, since it enables us to combine the learnt neural likelihood on small scales with the analytic likelihood on large scales. 

%%%%%%%%%%%%%%%%%%%%%%%%%%%%%%%%%%%%
\section{Hybrid Simulation-based Inference}
\label{sec:hysbi}

\noindent Consider the scenario in which the data-vector $\bx$ can be split into two components -- large scales $\bx_L$, and small scales $\bx_S$ -- such that $\bx=\{\bx_L, \bx_S\}$. Such a split is natural for the power spectrum (where $\bx_L$ would comprise measurements with wavenumbers below some transition scale, \textit{i.e.}\ $\{P(k): k\leq k_*\}$); however, we note that the following discussion is applicable to any combination of large- and small-scale statistics (e.g., power spectra on large scales and wavelets on small scales). The data likelihood can then be decomposed as the product of a large-scale likelihood and the \textit{conditional} likelihood of small-scales given large-scales:
\begin{equation}\label{eq: master-eq}
    p(\bx | \btheta) = p(\bx_L, \bx_S | \btheta) = p(\bx_L | \btheta) p(\bx_S | \bx_L, \btheta).
\end{equation}

In HySBI, we propose to use classical statistics such as the power spectrum and bispectrum on large scales for which the likelihood, $p(\bx_L | \btheta)$, can be modeled analytically with perturbation theory \cite[e.g.,][]{ivanov2020,Philcox:2021kcw,DAmico:2022osl}. As such, only the likelihood term on small scales, $p(\bx_S | \bx_L, \btheta)$, needs to be learnt with simulations.\footnote{This differs from previous hybrid approaches, such as Ref.\,\citep{Kokron:2021xgh, Modi2020}, who use small-volume simulations to model the signal, but assume a Gaussian likelihood; this fails for many higher-order statistics.} We stress that, in contrast to global SBI approaches, this likelihood term depends not only on the model parameters $\btheta$ but also on the large scale statistic $\bx_L$.

As discussed above, the main advantage of learning only the small-scale likelihood with simulations is that, in principle, it can be done by simulating only a small fractional volume at full fidelity instead of the whole simulation box, creating a surrogate likelihood that mitigates the increasing computational cost of scaling SBI to upcoming cosmological surveys. For a sufficiently large sub-volume, we still have enough modes on small scales that the associated sample variance is sub-dominant compared to other errors such as data noise and stochasticity in training.
% \oliver{I think we need to make this paragraph clearer, since it is a major part of this HySBI procedure. There's probably two issues: (a) simulating small-volumes with the known realizations of big boxes [which we bypass here], (b) super-sample variance, which is basically due to only looking at one (or few) subboxes.} 
%  However, this results in two new issues-
%  (1) this requires consistent simulation of small- and large-scale statistics, \textit{i.e.}\ one strictly needs to know the large-scale realization in which the small-scale Universe forms and use it to estimate $\bx_L$ correctly to learn the conditional dependence $p(\bx_S | \bx_L, \btheta)$; and (2) a given sub-volume suffers from super-sample effects from the large box that will impact the evolution in a given sub-volume, thus adding noise to the small scale statistics $\bx_S$ \oliver{add cites e.g. spergel}. 
% In this first work, we present a proof-of-principle of our methodology and study the impact of this super-sample variance. In the discussion below, we will comment on other subtleties associated with this approach.
 However, two new issues arise:
 (1) to correctly learn the conditional dependence $p(\bx_S | \bx_L, \btheta)$, one needs to be able to estimate the correct large-scale statistics $\bx_L$ for the exact large-scale realization in which the small-scale Universe forms with only coarse simulations on the full volume and full fidelity simulations only on the small volume; (2) a given sub-volume suffers from increased sample variance and super-sample effects from the large box that will impact the evolution in a given sub-volume, thus adding noise to the small scale statistics $\bx_S$ used for training SBI. 
 Thus we learn a \emph{surrogate likelihood} for the small scale data which has the same mean but larger volume and hence is more conservative (effectively trading bias from poor quality simulations with variance from a smaller simulation box).
In this first work, we present a proof-of-principle of our methodology and focus on studying the impact of this super-sample variance. In the discussion below, we will comment on other subtleties associated with this approach.

%%%%%%%%%%%%%%%%%%%%%%%%%%%%%%%%%%%%
\section{Data Analysis}
\label{sec:data}

\paragraph{Setup --}
In order to present our methodology and its advantages, this work considers constraining the $\Omega_m$ and $\sigma_8$ parameters from the three-dimensional dark matter distribution. Our motivation to use dark matter density field rather than a more realistic halo or galaxy sample for this first exercise is to avoid the need to infer non-cosmological parameters (such as those of the halo-occupation distribution), allowing us to focus solely on the impact of super-sample variance. 

To train the small-scale likelihood we use the {\sc quijote} latin-hypercube suite~\citep{villaescusa-navarro2020} which contains 2000 $N$-body simulations varying five cosmology parameters: $\{\Omega_{\mathrm m}, \, \sigma_{8}, \, \Omega_{\rm b}, \, n_s, h\}$. These simulations evolve $1024^3$ cold dark matter particles in a $1\,h^{-3}\mathrm{Gpc}^3$ box with the TreePM Gadget-III code \citep{Springel:2005mi}.
To mimic a small-volume simulation, we split each simulation into eight sub-volumes with side-length $500\,h^{-1}\mathrm{Mpc}$. 
1500 simulations are used for training, 200 for validation, and the remaining 300 for testing. 
We emphasize that we use the sub-volumes to estimate small scale statistics only for \emph{training HySBI} and in tests, we always analyze the statistics estimated from the whole-volume as simulated data. 

% \CM{This following section has gotten a bit clunky so can consider restructuring if it seems messy}
\vspace{5px}
\paragraph{Large-scale statistics --} Here, we set $\bx_L$ equal to the observed power spectrum of dark matter density field, $\hat{P}(k)$, on scales $k\in[0.007,0.15]\,h\mathrm{Mpc}^{-1}$, obtained from the full simulation volume of $1\,h^{-3}\mathrm{Gpc}^3$ with $\Delta k = 2\pi/1000\,h\,\mathrm{Mpc}^{-1}$, \textit{i.e.}\ the fundamental mode of the box.
This scale cut is chosen to balance two competing factors: (a) perturbation theory predictions are accurate only on large-scales (favoring lower $\kmax$); (b) to suppress sample-variance in the small-scale sub-box statistics, we require many modes and thus larger $\kmax$.
Power spectra are estimated with \textsc{nbodykit}, using cloud-in-cell interpolation on an $N=256$ side mesh \citep{hand2018}.

We adopt a Gaussian model for the large-scale likelihood, as specified by the Effective Field Theory of Large Scale Structure:
\beq
    -2\log p(\bx_L|\btheta) &=& \sum_k\left[\frac{P_{\rm loop}(k)-2c_s^2P_{\rm ct}(k)-\hat{P}(k)}{\sigma_P(k)}\right]^2,
\eeq
where the variance term is specified by the linear theory power spectrum, evaluated at the true cosmology of the test simulations. Our model for the theoretical power spectrum is given by 1-loop standard perturbation theory ($P_{\rm loop}(k)$) in addition to a renormalization contribution $-2c_s^2P_{\rm ct}(k)$, depending on the counter-term $c_s^2$, which must be treated as a free parameter and marginalized over \citep[e.g.,][]{Baumann:2010tm,Carrasco:2012cv}.\footnote{The PT models are computed via loop integrals over the linear power spectrum (with e.g., \textsc{class-pt} \citep{Chudaykin:2020aoj}), itself computed using Boltzmann solvers such as \textsc{class} or \textsc{camb}. For the sake of speed, here we instead train a simple neural network emulator to predict $P_{\mathrm{loop}}$ and $P_{\mathrm{ct}}$ as a function of cosmological parameters (given a grid of training data from \textsc{class-pt}). We have validated that this does not impact the final inference but speeds up our MCMC chains by a factor $\approx 100$.}

% \vspace{5px}
\paragraph{Small scale statistics --}
We consider two different choices for the small-scale statistic $\bx_S$: (1) the power spectrum for $k\leq 0.5\,h\,\mathrm{Mpc}^{-1}$; (2) wavelet coefficients.\footnote{Since the wavelets are not localized in Fourier space, it is hard to define a strict cut-off. The coefficients used here are constructed using   a mother wavelet whose weight falls to zero at the Nyquist frequency $k=0.8\,h\,\mathrm{Mpc}^{-1}$.} In both cases, the conditional likelihood model $p(\bx_S | \bx_L, \btheta)$, is learnt using SBI, given the above $\bx_L$ definition. To train the likelihood model, we estimate the small-scale statistics using only the sub-volumes of the simulation, one at a time; this emulates the realistic setting when it is practical only to compute reduced-volume simulations. We thus have eight estimates of $\bx_S$ from each simulation which generically differ due to sample and super-sample variance effects. To test our methodology, however, we generate $\bx_S$ from the full simulation volume of $1\,h^{-3}\mathrm{Gpc}^3$ such that all the observational data is used.

Here, we measure the power spectrum on scales $k\in [0.15,0.5]\,h\,\mathrm{Mpc}^{-1}$ with $\Delta k$ fixed to the fundamental mode of the small sub-volume, \textit{i.e.}\ $2\pi/500\,h\,\mathrm{Mpc}^{-1}$; this allows for sub-volume simulations and full-volume `data' to be robustly compared. Wavelet coefficients are defined for a statistically homogeneous real field $F$ as $S_0 = \langle |F|^p \rangle$ and $S_1 = \langle |I \ast \psi_j|^p \rangle$ where the dilation index $j$ varies as $q/Q$ with $q \in \{0, \dots, J\times Q-1\}$. The parameter $J$ is the total number of octaves \textit{i.e.}\ doublings in scale of the mother wavelet, and $Q$ is the quality factor, which controls the number of scales per octave. 
We use the package \texttt{galactic-wavelets}\footnote{\url{https://github.com/bregaldo/galactic\_wavelets}} to estimate these wavelet coefficients. Our mother wavelet is inspired from \cite{Lanusse2012} and we take $J=3$, $Q=4$, and $p\in\{0.5, 1, 1.5, 2\}$, resulting in a total of four $S_0$ and 48 $S_1$ wavelet coefficients \cite{Eickenberg2022}.

\paragraph{Non-periodic boundary conditions --}
Being cut-outs from a larger simulation, the sub-volumes do not have periodic boundary conditions, which must be accounted for when estimating the small-scale power spectrum and wavelet coefficients. 
For power spectrum, this is achieved via a (square) window-function matrix $M$ such that the true small scale power is $P(k) = [M^{-1} \tilde{P}](k)$, where $\tilde{P}$ is the measured small-scale power spectrum on the volume, assuming periodicity. This matrix is estimated by generating Gaussian white-noise fields over the full-volume with power injected only in a single $k-$bin, and then measuring the leakage of this power into other bins when power spectrum is evaluated on a sub-volume. 

Performing similar corrections for wavelet coefficients is easier since they are localized in configuration space. Here, we achieve this by masking the pixels near the boundary such that for any wavelet centered on a pixel within the mask, the fraction of weight that leaks outside the sub-volume is less than a pre-defined small threshold. This implies that larger the wavelets, the more pixels are masked. As such, the threshold determines the bias-variance trade-off in our estimator, and is here set to 0.02.

\paragraph{SBI Implementation --}
We use the \texttt{sbi}\footnote{\url{https://github.com/mackelab/sbi}} package to train masked auto-regressive flows
as conditional neural density estimators and learn the likelihood, $q_{\phi}(\bx_S | \bx_L, \btheta) \sim p(\bx_S | \bx_L, \btheta)$.
To minimize stochasticity in training, we train 400 networks for each statistic by varying hyperparameters such as the network width, number of layers, learning rate and batch size.
We use the \texttt{Weights-and-Biases}\footnote{\url{https://wandb.ai/site}} package for the hyperparameter search.
After training, we collect the ten neural density estimators with best validation loss and use them as an ensemble.

\paragraph{Posterior Inference --}
For inference over a new observation $\bx'$, we combine the learnt small-scale likelihood with the Gaussian likelihood for large scale $p(\bx_L | \btheta)$ (via Eq.\,\ref{eq: master-eq}), and a uniform prior on cosmology parameters (dictated by the bounds of the \textsc{quijote} latin-hypercube). The counter-term $c_s$ has a Gaussian prior $\mathcal{N}(0,10)$ \citep[e.g.,][]{Philcox:2021kcw}. To generate samples from the posterior, we use the \texttt{emcee}  package \citep{Foreman-Mackey:2012any} for MCMC sampling and run 20 chains with 11,000 samples each.  

\section{Results}
\label{sec:results}

\begin{figure}
\centering
\includegraphics[width=0.98\columnwidth]{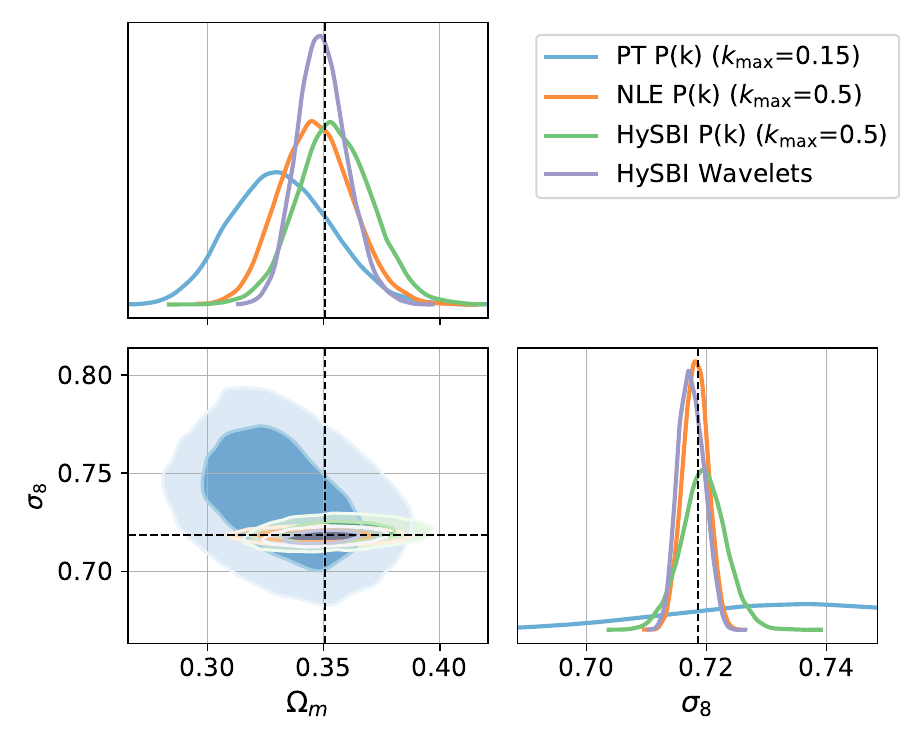}
\caption{Posterior distribution of $\Omega_m$ and $\sigma_8$ inferred from a power spectrum analysis using classical perturbative methods on large-scales (PT), full simulation-based inference (NLE), and our new hybrid scheme (HySBI). We also show HySBI results including small-scale wavelet statistics (purple). Results are shown for a representative test simulation (\textsc{quiojte} LH-872). HySBI strongly outperforms classical techniques, and is able to robustly infer cosmological parameters without large-volume simulations.
%Posterior distribution of $\Omega_m$ and $\sigma_8$ for a test simulation, LH 872.
%We compare the results for perturbation theory, SBI (NLE) with power spectrum over the full range of scales, and HySBI combining PT on large scales with SBI for P(k) and wavelet coefficients on small scales.
}
\label{fig:moneyplot}
\end{figure}

\noindent In Fig.~\ref{fig:moneyplot}, we show the results of HySBI using the power spectrum and wavelet coefficients for inferring $\Omega_m$ and $\sigma_8$ on one of the test simulations\footnote{Results are similar on all the 160 simulations in the test dataset.}.
To minimize the impact of super-sample variance, the SBI component of HySBI is trained using all the eight sub-volumes of the simulations \textit{i.e.}\ we take the mean of the small-scale statistics estimated from all the individual sub-volumes for training.\footnote{The mean of all subvolume statistics well approximates the statistic computed on the total volume; differences appear only from boundary effects, which are small if the characteristic scale of $\bx$ is small compared to the subbox size.} As discussed in the previous section, the test data is generated from the full $1\,h^{-3}\mathrm{Gpc}^3$ simulation volume. 

For comparison, we show also the results of the PT analysis which is restricted to the large scales ($k \in [0.007, 0.15]\,\ihmpc$), and a global SBI power spectrum analysis (hereafter NLE) that covers the same scales as HySBI ($k \in [0.007, 0.5]\,\ihmpc$). As expected, HySBI significantly outperforms the classical PT analysis, clearly demonstrating the utility of our method. That said, the NLE power spectrum constraints on $\sigma_8$ are somewhat tighter than those of HySBI over the same range; this is because, in our scale-decoupled formalism, HySBI also infers and marginalizes over the large-scale PT counter-term $c_s^2$, unlike global SBI.
This can in principle be improved upon by learning a prior of $c_s^2$ conditioned on small scale statistics and cosmology, but we do not explore that in this work. Finally, HySBI with wavelet coefficients on small scales outperforms all the power spectrum analyses, emphasizing the advantage of using higher order statistics.

\begin{figure}
\centering
\includegraphics[width=0.98\columnwidth]{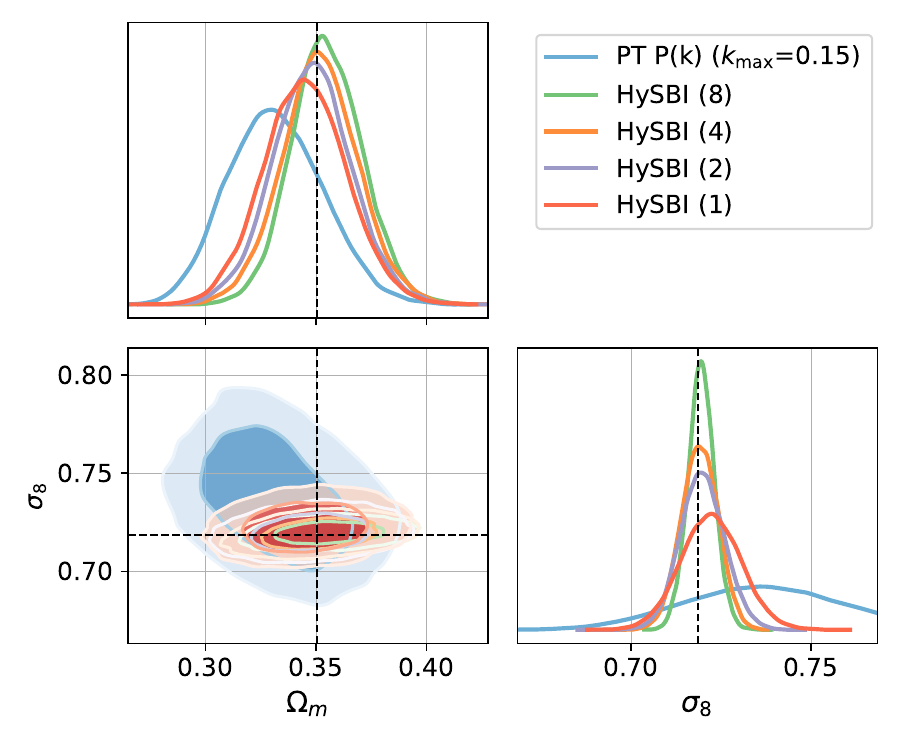}
\caption{
As Fig.\,\ref{fig:moneyplot}, but comparing the power spectrum HySBI constraints when the SBI is trained using one, two, four, and eight sub-volumes of every simulation. This reduces super-sample variance in the small-scale training data, and is found to somewhat enhance $\sigma_8$ constraints.}
\label{fig:pk_nsubs}
\end{figure}

\begin{figure}
\centering
\includegraphics[width=0.98\columnwidth]{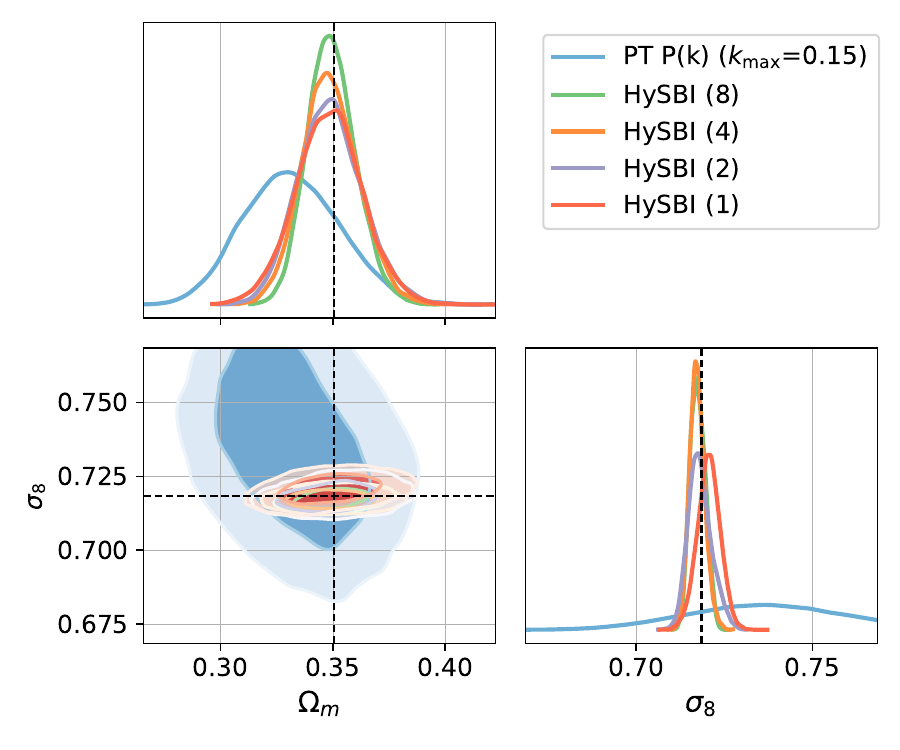}
\caption{As Fig.\,\ref{fig:pk_nsubs}, but using wavelet coefficients as the small-scale statistic. The improvement in $\sigma_8$ with number of sub-volumes is more muted in this case. 
%Posteriors distribution of $\Omega_m$ and $\sigma_8$ for LH 872 when HySBI with wavelet coefficients on small scales is trained using 1, 2, 4 and 8 sub-volumes of every simulation. 
}
\label{fig:wv_nsubs}
\end{figure}

To study the impact of super-sample covariance, we repeat the HySBI analysis training the small-scale likelihood using different number of sub-volumes (each of size $500^3\,h^{-3}\mathrm{Mpc}^3$). The corresponding constraints on $\Omega_m$ and $\sigma_8$ are shown in Fig.~\ref{fig:pk_nsubs} for a single representative simulation using the power spectrum as the summary statistic for both small and large scales. Averaging over all test simulations, the uncertainties when using four, two and one sub-volumes are inflated by $\sim 6,\,9,\,11\%$ ($\sim 45,\, 80,\, 130\%$) for $\Omega_m$ ($\sigma_8$) compared to those using all eight sub-volumes. The inference of $\sigma_8$ is thus more sensitive to super-sample effects than $\Omega_m$; this matches expectations, since most of the constraining power on the latter comes from the large scales which are modeled with PT, whilst $\sigma_8$ constraints are most informed by small scales. 

In Fig.~\ref{fig:wv_nsubs}, we show the same analysis using wavelet coefficients on small scales. In this case, we find that constraints are inflated by $\sim 20,\,24,\,50 \%$ ($\sim 40,\, 80,\, 100\%$) for $\Omega_m$ ($\sigma_8$); here the $\Omega_m$ analysis more sensitive to super-sample effects than before, since the wavelets carry significant information on the matter density on all scales. On the other hand, $\sigma_8$ constraints are slightly less sensitive for wavelet coefficients than the power spectrum since wavelets include statistics corresponding to $\delta^p$ for $p=0.5,1,1.5$ (where $\delta$ is the overdensity field), whilst the power spectrum only measures squared statistics, \textit{i.e.}\ $\delta^2$. The former lead to lower sample variance across the different sub-volumes.

\section{Discussion and Outlook}
\label{sec:discussion}

\noindent 
% The past iteration of cosmological data has been analyzed primarily through traditional analytical methods such as perturbation theory (PT); whilst this has been highly successful \oliver{cite}, the approach is necessarily limited to low-order clustering statistics (such as the two- and three-point functions), and to linear and quasi-linear scales. If we wish to fully exploit the information contained within upcoming large scale structure surveys, new methods are needed. 

Simulation-based inference provides a promising avenue through which to overcome the limitations of traditional analytic methods for cosmological analysis, and make use of the upcoming tranche of small-scale data. However, the computational cost of generating the high-fidelity and high-volume simulations needed to train SBI at the precision required for the next generation of cosmological surveys will likely be prohibitive. 

To mitigate this challenge, we propose Hybrid simulation-based inference (HySBI); this combines perturbative methods on large scales (using quasi-optimal statistics such as the power spectrum and bispectrum), with simulation-based techniques on small scales (including arbitrary higher-order statistics which cannot be modeled analytically). This is achieved by decomposing the data likelihood as the product of a large-scale likelihood modeled analytically with perturbation theory, and a \textit{conditional} likelihood of small-scales given large-scales -- the latter is learnt using simulations. This approach can be lead to significant computational gains, since we need to simulate only a small fraction of the whole data volume to model the whole.

A natural drawback is that modeling small-scales using only one (or a few) sub-volumes of a simulation adds super-sample effects from the large scale modes in the global simulation box.
As a result, we learn a conservative surrogate likelihood for small scales which has a larger variance (but removes the bias that would result from simulating the entire volume at low resolution). In this first work, we have presented a proof-of-principle of HySBI and studied the impact of this super-sample variance by inferring $\sigma_8$ and $\Omega_m$ using the three-dimensional distribution of the dark-matter density field, analyzed with the large-scale power spectrum, and small-scale power spectrum or wavelet coefficients. Upon splitting the entire $1 h^{-3}\mathrm{Gpc}^3$ simulation box into eight sub-volumes, we found that using even one of the eight sub-volumes to train the small-scale likelihood leads to significant gains over state-of-the-art power spectrum analyses.

Performing a global SBI analysis leads to slightly better results than our hybrid scheme; this occurs since we have enforced that the large-scale likelihood is not influenced by small-scales, and thus marginalized over a necessary perturbative counterterm parameter appearing on large scales. Such an assumption must be altered for realistic observables such as galaxies, whence there are nuisance parameters in both PT and SBI (encoding bias and halo-occupation distributions). These parameters however are not independent and should be inferred jointly. This will be explored in future work.

To fully realize the computational advanced promised by HySBI, two technical advances must be realized. As mentioned briefly above, a key ingredient of our SBI framework is accurate, high-fidelity small sub-volume training simulations which ``know'' about the large-volume Universe in which they are embedded, \textit{i.e.}\ those from which we can measure both the small-scale statistics, $\bx_S$, and the corresponding realization of the large-scale data, $\bx_L$ but without running the full-fidelity simulations on the full volume for the latter.  We plan on developing these simulations by building upon promising techniques that are already available. For example, the s-COLA method creates coarse PT simulations on large-scales (which provide the global $\bx_L$ measurements), and evolves small sub-volumes with high-resolution particle-mesh simulations \citep{Tassev:2015mia, Leclercq:2020nmz}. The FlowPM simulations, which use multi-grid schemes to estimate small-scale forces independently in different sub-volumes, % and combine them with globally estimated large scale forces, 
can also be adapted to this end \citep{modi2021}. A third avenue is by building upon the zoom-in simulations for galaxy formation \citep{Nadler2023}. 
Crucially, sub-volumes can be independently simulated in these frameworks, thus we can straightforwardly run a single small-volume simulation with a known (coarsened) large-volume realization. Since these can be run on a single GPU, the resulting pipeline will not require multi-node communications and can enjoy GPU accelerations; this will lead to significant computational gains even if we simulate all the sub-volumes of a given simulation to minimize super-sample effects \citep{Li:2022qlf,modi2021}. 

A second technical hurdle to be overcome is the the inclusion of systematic effects such as survey masks, fibre collisions \textit{et cetera}. These often couple small- and large-scales and will be considered in future works. Finally, we currently assume a clear division between the large and small scale statistics (such that $\bx_S$ can be measured from sub-boxes), which excludes some types of statistics such as squeezed bispectra; similar techniques can likely be developed for such modified analyses (often of use in inflationary studies) \citep[e.g.,][]{Goldstein:2022hgr,Cabass:2022wjy,Cabass:2022ymb,DAmico:2022gki}.%However we believe that it is a minor shortcoming and overall does not take away from the potential gains of being able to realistically scale simulation-based inference techniques for analyzing the next generation of cosmological surveys. 
Though much work remains to be done, it seems likely that hybrid approaches such as HySBI can create a promising symbiosis of old and new techniques, significantly enhancing the science output of future surveys without requiring prohibitive computational costs.

%\newpage
\vskip 8pt
\begin{acknowledgments}
\footnotesize{
\begingroup
\hypersetup{hidelinks}
OHEP is a Junior Fellow of the Simons Society of Fellows and thanks the office of Chuck Schumer for immigration assistance. CM would like to thank \href{https://www.imdb.com/title/tt0057012/}{Dr.\,Strangelove} for the title, and Ben Wandelt, Michael Eickenberg, and Mikhail Ivanov for useful discussions. We would especially like to thank Bruno Régaldo-Saint Blancard for setting up and sharing the code for estimating wavelet coefficients with erosion. We additionally thank Sebastian Wagner-Carena for insightful comments on the draft. This work was partly conceived during the ``Beyond 2-point Challenge'' workshop at the University of Arizona in February 2023. 
\endgroup
}
\end{acknowledgments}

\appendix

% \section{Appendixes}


% \nocite{*}

\bibliography{biblio}% Produces the bibliography via BibTeX.

\begin{thebibliography}{38}%
\makeatletter
\providecommand \@ifxundefined [1]{%
 \@ifx{#1\undefined}
}%
\providecommand \@ifnum [1]{%
 \ifnum #1\expandafter \@firstoftwo
 \else \expandafter \@secondoftwo
 \fi
}%
\providecommand \@ifx [1]{%
 \ifx #1\expandafter \@firstoftwo
 \else \expandafter \@secondoftwo
 \fi
}%
\providecommand \natexlab [1]{#1}%
\providecommand \enquote  [1]{``#1''}%
\providecommand \bibnamefont  [1]{#1}%
\providecommand \bibfnamefont [1]{#1}%
\providecommand \citenamefont [1]{#1}%
\providecommand \href@noop [0]{\@secondoftwo}%
\providecommand \href [0]{\begingroup \@sanitize@url \@href}%
\providecommand \@href[1]{\@@startlink{#1}\@@href}%
\providecommand \@@href[1]{\endgroup#1\@@endlink}%
\providecommand \@sanitize@url [0]{\catcode `\\12\catcode `\$12\catcode
  `\&12\catcode `\#12\catcode `\^12\catcode `\_12\catcode `\%12\relax}%
\providecommand \@@startlink[1]{}%
\providecommand \@@endlink[0]{}%
\providecommand \url  [0]{\begingroup\@sanitize@url \@url }%
\providecommand \@url [1]{\endgroup\@href {#1}{\urlprefix }}%
\providecommand \urlprefix  [0]{URL }%
\providecommand \Eprint [0]{\href }%
\providecommand \doibase [0]{https://doi.org/}%
\providecommand \selectlanguage [0]{\@gobble}%
\providecommand \bibinfo  [0]{\@secondoftwo}%
\providecommand \bibfield  [0]{\@secondoftwo}%
\providecommand \translation [1]{[#1]}%
\providecommand \BibitemOpen [0]{}%
\providecommand \bibitemStop [0]{}%
\providecommand \bibitemNoStop [0]{.\EOS\space}%
\providecommand \EOS [0]{\spacefactor3000\relax}%
\providecommand \BibitemShut  [1]{\csname bibitem#1\endcsname}%
\let\auto@bib@innerbib\@empty
%</preamble>
\bibitem [{\citenamefont {Collaboration}\ \emph
  {et~al.}(2016{\natexlab{a}})\citenamefont {Collaboration}, \citenamefont
  {Aghamousa}, \citenamefont {Aguilar}, \citenamefont {Ahlen}, \citenamefont
  {Alam}, \citenamefont {Allen}, \citenamefont {Prieto}, \citenamefont {Annis},
  \citenamefont {Bailey}, \citenamefont {Balland} \emph
  {et~al.}}]{desicollaboration2016}%
  \BibitemOpen
  \bibfield  {author} {\bibinfo {author} {\bibfnamefont {D.}~\bibnamefont
  {Collaboration}}, \bibinfo {author} {\bibfnamefont {A.}~\bibnamefont
  {Aghamousa}}, \bibinfo {author} {\bibfnamefont {J.}~\bibnamefont {Aguilar}},
  \bibinfo {author} {\bibfnamefont {S.}~\bibnamefont {Ahlen}}, \bibinfo
  {author} {\bibfnamefont {S.}~\bibnamefont {Alam}}, \bibinfo {author}
  {\bibfnamefont {L.~E.}\ \bibnamefont {Allen}}, \bibinfo {author}
  {\bibfnamefont {C.~A.}\ \bibnamefont {Prieto}}, \bibinfo {author}
  {\bibfnamefont {J.}~\bibnamefont {Annis}}, \bibinfo {author} {\bibfnamefont
  {S.}~\bibnamefont {Bailey}}, \bibinfo {author} {\bibfnamefont
  {C.}~\bibnamefont {Balland}}, \emph {et~al.},\ }\bibfield  {title} {\bibinfo
  {title} {The {{DESI Experiment Part I}}: {{Science}},{{Targeting}}, and
  {{Survey Design}}},\ }\href@noop {} {\bibfield  {journal} {\bibinfo
  {journal} {arXiv:1611.00036 [astro-ph]}\ } (\bibinfo {year}
  {2016}{\natexlab{a}})},\ \Eprint {https://arxiv.org/abs/1611.00036}
  {arXiv:1611.00036 [astro-ph]} \BibitemShut {NoStop}%
\bibitem [{\citenamefont {Collaboration}\ \emph
  {et~al.}(2016{\natexlab{b}})\citenamefont {Collaboration}, \citenamefont
  {Aghamousa}, \citenamefont {Aguilar}, \citenamefont {Ahlen}, \citenamefont
  {Alam}, \citenamefont {Allen}, \citenamefont {Prieto}, \citenamefont {Annis},
  \citenamefont {Bailey}, \citenamefont {Balland}, \citenamefont {Ballester}
  \emph {et~al.}}]{desicollaboration2016a}%
  \BibitemOpen
  \bibfield  {author} {\bibinfo {author} {\bibfnamefont {D.}~\bibnamefont
  {Collaboration}}, \bibinfo {author} {\bibfnamefont {A.}~\bibnamefont
  {Aghamousa}}, \bibinfo {author} {\bibfnamefont {J.}~\bibnamefont {Aguilar}},
  \bibinfo {author} {\bibfnamefont {S.}~\bibnamefont {Ahlen}}, \bibinfo
  {author} {\bibfnamefont {S.}~\bibnamefont {Alam}}, \bibinfo {author}
  {\bibfnamefont {L.~E.}\ \bibnamefont {Allen}}, \bibinfo {author}
  {\bibfnamefont {C.~A.}\ \bibnamefont {Prieto}}, \bibinfo {author}
  {\bibfnamefont {J.}~\bibnamefont {Annis}}, \bibinfo {author} {\bibfnamefont
  {S.}~\bibnamefont {Bailey}}, \bibinfo {author} {\bibfnamefont
  {C.}~\bibnamefont {Balland}}, \bibinfo {author} {\bibfnamefont
  {O.}~\bibnamefont {Ballester}}, \emph {et~al.},\ }\bibfield  {title}
  {\bibinfo {title} {The {{DESI Experiment Part II}}: {{Instrument Design}}},\
  }\href@noop {} {\bibfield  {journal} {\bibinfo  {journal} {arXiv:1611.00037
  [astro-ph]}\ } (\bibinfo {year} {2016}{\natexlab{b}})},\ \Eprint
  {https://arxiv.org/abs/1611.00037} {arXiv:1611.00037 [astro-ph]} \BibitemShut
  {NoStop}%
\bibitem [{\citenamefont {Abareshi}\ \emph {et~al.}(2022)\citenamefont
  {Abareshi}, \citenamefont {Aguilar}, \citenamefont {Ahlen}, \citenamefont
  {Alam}, \citenamefont {Alexander}, \citenamefont {Alfarsy}, \citenamefont
  {Allen}, \citenamefont {Prieto} \emph {et~al.}}]{abareshi2022}%
  \BibitemOpen
  \bibfield  {author} {\bibinfo {author} {\bibfnamefont {B.}~\bibnamefont
  {Abareshi}}, \bibinfo {author} {\bibfnamefont {J.}~\bibnamefont {Aguilar}},
  \bibinfo {author} {\bibfnamefont {S.}~\bibnamefont {Ahlen}}, \bibinfo
  {author} {\bibfnamefont {S.}~\bibnamefont {Alam}}, \bibinfo {author}
  {\bibfnamefont {D.~M.}\ \bibnamefont {Alexander}}, \bibinfo {author}
  {\bibfnamefont {R.}~\bibnamefont {Alfarsy}}, \bibinfo {author} {\bibfnamefont
  {L.}~\bibnamefont {Allen}}, \bibinfo {author} {\bibfnamefont {C.~A.}\
  \bibnamefont {Prieto}}, \emph {et~al.},\ }\href
  {https://doi.org/10.48550/arXiv.2205.10939} {\bibinfo {title} {Overview of
  the {{Instrumentation}} for the {{Dark Energy Spectroscopic Instrument}}}}
  (\bibinfo {year} {2022}),\ \Eprint {https://arxiv.org/abs/2205.10939}
  {arXiv:2205.10939 [astro-ph]} \BibitemShut {NoStop}%
\bibitem [{\citenamefont {Laureijs}\ \emph {et~al.}(2011)\citenamefont
  {Laureijs}, \citenamefont {Amiaux}, \citenamefont {Arduini}, \citenamefont
  {Augu{\`e}res}, \citenamefont {Brinchmann}, \citenamefont {Cole},
  \citenamefont {Cropper}, \citenamefont {Dabin}, \citenamefont {Duvet},
  \citenamefont {Ealet} \emph {et~al.}}]{laureijs2011}%
  \BibitemOpen
  \bibfield  {author} {\bibinfo {author} {\bibfnamefont {R.}~\bibnamefont
  {Laureijs}}, \bibinfo {author} {\bibfnamefont {J.}~\bibnamefont {Amiaux}},
  \bibinfo {author} {\bibfnamefont {S.}~\bibnamefont {Arduini}}, \bibinfo
  {author} {\bibfnamefont {J.-L.}\ \bibnamefont {Augu{\`e}res}}, \bibinfo
  {author} {\bibfnamefont {J.}~\bibnamefont {Brinchmann}}, \bibinfo {author}
  {\bibfnamefont {R.}~\bibnamefont {Cole}}, \bibinfo {author} {\bibfnamefont
  {M.}~\bibnamefont {Cropper}}, \bibinfo {author} {\bibfnamefont
  {C.}~\bibnamefont {Dabin}}, \bibinfo {author} {\bibfnamefont
  {L.}~\bibnamefont {Duvet}}, \bibinfo {author} {\bibfnamefont
  {A.}~\bibnamefont {Ealet}}, \emph {et~al.},\ }\bibfield  {title} {\bibinfo
  {title} {Euclid {{Definition Study Report}}},\ }\href@noop {} {\bibfield
  {journal} {\bibinfo  {journal} {arXiv e-prints}\ ,\ \bibinfo {pages}
  {arXiv:1110.3193}} (\bibinfo {year} {2011})}\BibitemShut {NoStop}%
\bibitem [{\citenamefont {{Ivezi{\'c}}}\ \emph {et~al.}(2019)\citenamefont
  {{Ivezi{\'c}}}, \citenamefont {{Kahn}}, \citenamefont {{Tyson}},
  \citenamefont {{Abel}}, \citenamefont {{Acosta}}, \citenamefont {{Allsman}},
  \citenamefont {{Alonso}}, \citenamefont {{AlSayyad}}, \citenamefont
  {{Anderson}}, \citenamefont {{Andrew}},\ and\ \citenamefont {et~al.}}]{lsst}%
  \BibitemOpen
  \bibfield  {author} {\bibinfo {author} {\bibfnamefont {{\v Z}.}~\bibnamefont
  {{Ivezi{\'c}}}}, \bibinfo {author} {\bibfnamefont {S.~M.}\ \bibnamefont
  {{Kahn}}}, \bibinfo {author} {\bibfnamefont {J.~A.}\ \bibnamefont {{Tyson}}},
  \bibinfo {author} {\bibfnamefont {B.}~\bibnamefont {{Abel}}}, \bibinfo
  {author} {\bibfnamefont {E.}~\bibnamefont {{Acosta}}}, \bibinfo {author}
  {\bibfnamefont {R.}~\bibnamefont {{Allsman}}}, \bibinfo {author}
  {\bibfnamefont {D.}~\bibnamefont {{Alonso}}}, \bibinfo {author}
  {\bibfnamefont {Y.}~\bibnamefont {{AlSayyad}}}, \bibinfo {author}
  {\bibfnamefont {S.~F.}\ \bibnamefont {{Anderson}}}, \bibinfo {author}
  {\bibfnamefont {J.}~\bibnamefont {{Andrew}}},\ and\ \bibinfo {author}
  {\bibnamefont {et~al.}},\ }\bibfield  {title} {\bibinfo {title} {{LSST: From
  Science Drivers to Reference Design and Anticipated Data Products}},\ }\href
  {https://doi.org/10.3847/1538-4357/ab042c} {\bibfield  {journal} {\bibinfo
  {journal} {\apj}\ }\textbf {\bibinfo {volume} {873}},\ \bibinfo {eid} {111}
  (\bibinfo {year} {2019})},\ \Eprint {https://arxiv.org/abs/0805.2366}
  {arXiv:0805.2366} \BibitemShut {NoStop}%
\bibitem [{\citenamefont {Spergel}\ \emph {et~al.}(2015)\citenamefont
  {Spergel}, \citenamefont {Gehrels}, \citenamefont {Baltay}, \citenamefont
  {Bennett}, \citenamefont {Breckinridge}, \citenamefont {Donahue},
  \citenamefont {Dressler}, \citenamefont {Gaudi}, \citenamefont {Greene},
  \citenamefont {Guyon} \emph {et~al.}}]{spergel2015}%
  \BibitemOpen
  \bibfield  {author} {\bibinfo {author} {\bibfnamefont {D.}~\bibnamefont
  {Spergel}}, \bibinfo {author} {\bibfnamefont {N.}~\bibnamefont {Gehrels}},
  \bibinfo {author} {\bibfnamefont {C.}~\bibnamefont {Baltay}}, \bibinfo
  {author} {\bibfnamefont {D.}~\bibnamefont {Bennett}}, \bibinfo {author}
  {\bibfnamefont {J.}~\bibnamefont {Breckinridge}}, \bibinfo {author}
  {\bibfnamefont {M.}~\bibnamefont {Donahue}}, \bibinfo {author} {\bibfnamefont
  {A.}~\bibnamefont {Dressler}}, \bibinfo {author} {\bibfnamefont {B.~S.}\
  \bibnamefont {Gaudi}}, \bibinfo {author} {\bibfnamefont {T.}~\bibnamefont
  {Greene}}, \bibinfo {author} {\bibfnamefont {O.}~\bibnamefont {Guyon}}, \emph
  {et~al.},\ }\href@noop {} {\bibinfo {title} {Wide-{{Field InfrarRed Survey
  Telescope-Astrophysics Focused Telescope Assets WFIRST-AFTA}} 2015
  {{Report}}}} (\bibinfo {year} {2015})\BibitemShut {NoStop}%
\bibitem [{\citenamefont {Wang}\ \emph {et~al.}(2022)\citenamefont {Wang},
  \citenamefont {Zhai}, \citenamefont {Alavi}, \citenamefont {Massara},
  \citenamefont {Pisani}, \citenamefont {Benson}, \citenamefont {Hirata},
  \citenamefont {Samushia}, \citenamefont {Weinberg} \emph
  {et~al.}}]{wang2022a}%
  \BibitemOpen
  \bibfield  {author} {\bibinfo {author} {\bibfnamefont {Y.}~\bibnamefont
  {Wang}}, \bibinfo {author} {\bibfnamefont {Z.}~\bibnamefont {Zhai}}, \bibinfo
  {author} {\bibfnamefont {A.}~\bibnamefont {Alavi}}, \bibinfo {author}
  {\bibfnamefont {E.}~\bibnamefont {Massara}}, \bibinfo {author} {\bibfnamefont
  {A.}~\bibnamefont {Pisani}}, \bibinfo {author} {\bibfnamefont
  {A.}~\bibnamefont {Benson}}, \bibinfo {author} {\bibfnamefont {C.~M.}\
  \bibnamefont {Hirata}}, \bibinfo {author} {\bibfnamefont {L.}~\bibnamefont
  {Samushia}}, \bibinfo {author} {\bibfnamefont {D.~H.}\ \bibnamefont
  {Weinberg}}, \emph {et~al.},\ }\bibfield  {title} {\bibinfo {title} {The
  {{High Latitude Spectroscopic Survey}} on the {{Nancy Grace Roman Space
  Telescope}}},\ }\href {https://doi.org/10.3847/1538-4357/ac4973} {\bibfield
  {journal} {\bibinfo  {journal} {The Astrophysical Journal}\ }\textbf
  {\bibinfo {volume} {928}},\ \bibinfo {pages} {1} (\bibinfo {year}
  {2022})}\BibitemShut {NoStop}%
\bibitem [{\citenamefont {Philcox}\ and\ \citenamefont
  {Ivanov}(2022)}]{Philcox:2021kcw}%
  \BibitemOpen
  \bibfield  {author} {\bibinfo {author} {\bibfnamefont {O.~H.~E.}\
  \bibnamefont {Philcox}}\ and\ \bibinfo {author} {\bibfnamefont {M.~M.}\
  \bibnamefont {Ivanov}},\ }\bibfield  {title} {\bibinfo {title} {{BOSS DR12
  full-shape cosmology: \ensuremath{\Lambda}CDM constraints from the
  large-scale galaxy power spectrum and bispectrum monopole}},\ }\href
  {https://doi.org/10.1103/PhysRevD.105.043517} {\bibfield  {journal} {\bibinfo
   {journal} {Phys. Rev. D}\ }\textbf {\bibinfo {volume} {105}},\ \bibinfo
  {pages} {043517} (\bibinfo {year} {2022})},\ \Eprint
  {https://arxiv.org/abs/2112.04515} {arXiv:2112.04515 [astro-ph.CO]}
  \BibitemShut {NoStop}%
\bibitem [{\citenamefont {D'Amico}\ \emph
  {et~al.}(2022{\natexlab{a}})\citenamefont {D'Amico}, \citenamefont {Donath},
  \citenamefont {Lewandowski}, \citenamefont {Senatore},\ and\ \citenamefont
  {Zhang}}]{DAmico:2022osl}%
  \BibitemOpen
  \bibfield  {author} {\bibinfo {author} {\bibfnamefont {G.}~\bibnamefont
  {D'Amico}}, \bibinfo {author} {\bibfnamefont {Y.}~\bibnamefont {Donath}},
  \bibinfo {author} {\bibfnamefont {M.}~\bibnamefont {Lewandowski}}, \bibinfo
  {author} {\bibfnamefont {L.}~\bibnamefont {Senatore}},\ and\ \bibinfo
  {author} {\bibfnamefont {P.}~\bibnamefont {Zhang}},\ }\bibfield  {title}
  {\bibinfo {title} {{The BOSS bispectrum analysis at one loop from the
  Effective Field Theory of Large-Scale Structure}},\ }\href@noop {} {\
  (\bibinfo {year} {2022}{\natexlab{a}})},\ \Eprint
  {https://arxiv.org/abs/2206.08327} {arXiv:2206.08327 [astro-ph.CO]}
  \BibitemShut {NoStop}%
\bibitem [{\citenamefont {Chen}\ \emph {et~al.}(2022)\citenamefont {Chen},
  \citenamefont {Vlah},\ and\ \citenamefont {White}}]{Chen:2021wdi}%
  \BibitemOpen
  \bibfield  {author} {\bibinfo {author} {\bibfnamefont {S.-F.}\ \bibnamefont
  {Chen}}, \bibinfo {author} {\bibfnamefont {Z.}~\bibnamefont {Vlah}},\ and\
  \bibinfo {author} {\bibfnamefont {M.}~\bibnamefont {White}},\ }\bibfield
  {title} {\bibinfo {title} {{A new analysis of galaxy 2-point functions in the
  BOSS survey, including full-shape information and post-reconstruction BAO}},\
  }\href {https://doi.org/10.1088/1475-7516/2022/02/008} {\bibfield  {journal}
  {\bibinfo  {journal} {JCAP}\ }\textbf {\bibinfo {volume} {02}}\bibfield
  {number} {\bibinfo  {number} { (02)},\ \bibinfo {pages} {008}},\ }\Eprint
  {https://arxiv.org/abs/2110.05530} {arXiv:2110.05530 [astro-ph.CO]}
  \BibitemShut {NoStop}%
\bibitem [{\citenamefont {Alsing}\ \emph {et~al.}(2018)\citenamefont {Alsing},
  \citenamefont {Wandelt},\ and\ \citenamefont {Feeney}}]{alsing2018}%
  \BibitemOpen
  \bibfield  {author} {\bibinfo {author} {\bibfnamefont {J.}~\bibnamefont
  {Alsing}}, \bibinfo {author} {\bibfnamefont {B.}~\bibnamefont {Wandelt}},\
  and\ \bibinfo {author} {\bibfnamefont {S.}~\bibnamefont {Feeney}},\
  }\bibfield  {title} {\bibinfo {title} {Massive optimal data compression and
  density estimation for scalable, likelihood-free inference in cosmology},\
  }\href@noop {} {\bibfield  {journal} {\bibinfo  {journal} {arXiv:1801.01497
  [astro-ph]}\ } (\bibinfo {year} {2018})},\ \Eprint
  {https://arxiv.org/abs/1801.01497} {arXiv:1801.01497 [astro-ph]} \BibitemShut
  {NoStop}%
\bibitem [{\citenamefont {Alsing}\ \emph {et~al.}(2019)\citenamefont {Alsing},
  \citenamefont {Charnock}, \citenamefont {Feeney},\ and\ \citenamefont
  {Wandelt}}]{alsing2019}%
  \BibitemOpen
  \bibfield  {author} {\bibinfo {author} {\bibfnamefont {J.}~\bibnamefont
  {Alsing}}, \bibinfo {author} {\bibfnamefont {T.}~\bibnamefont {Charnock}},
  \bibinfo {author} {\bibfnamefont {S.}~\bibnamefont {Feeney}},\ and\ \bibinfo
  {author} {\bibfnamefont {B.}~\bibnamefont {Wandelt}},\ }\bibfield  {title}
  {\bibinfo {title} {Fast likelihood-free cosmology with neural density
  estimators and active learning},\ }\href
  {https://doi.org/10.1093/mnras/stz1960} {\bibfield  {journal} {\bibinfo
  {journal} {Monthly Notices of the Royal Astronomical Society}\ }\textbf
  {\bibinfo {volume} {488}},\ \bibinfo {pages} {4440} (\bibinfo {year}
  {2019})}\BibitemShut {NoStop}%
\bibitem [{\citenamefont {Jeffrey}\ \emph {et~al.}(2021)\citenamefont
  {Jeffrey}, \citenamefont {Alsing},\ and\ \citenamefont
  {Lanusse}}]{jeffrey2021}%
  \BibitemOpen
  \bibfield  {author} {\bibinfo {author} {\bibfnamefont {N.}~\bibnamefont
  {Jeffrey}}, \bibinfo {author} {\bibfnamefont {J.}~\bibnamefont {Alsing}},\
  and\ \bibinfo {author} {\bibfnamefont {F.}~\bibnamefont {Lanusse}},\
  }\bibfield  {title} {\bibinfo {title} {Likelihood-free inference with neural
  compression of {{DES SV}} weak lensing map statistics},\ }\href
  {https://doi.org/10.1093/mnras/staa3594} {\bibfield  {journal} {\bibinfo
  {journal} {Monthly Notices of the Royal Astronomical Society}\ }\textbf
  {\bibinfo {volume} {501}},\ \bibinfo {pages} {954} (\bibinfo {year}
  {2021})}\BibitemShut {NoStop}%
\bibitem [{\citenamefont {{Hahn}}\ \emph {et~al.}(2022)\citenamefont {{Hahn}},
  \citenamefont {{Eickenberg}}, \citenamefont {{Ho}}, \citenamefont {{Hou}},
  \citenamefont {{Lemos}}, \citenamefont {{Massara}}, \citenamefont
  {{Modi}~Chirag}, \citenamefont {{R\'egaldo-Saint Blancard}},\ and\
  \citenamefont {{Abidi}}}]{simbigletter}%
  \BibitemOpen
  \bibfield  {author} {\bibinfo {author} {\bibfnamefont {C.}~\bibnamefont
  {{Hahn}}}, \bibinfo {author} {\bibfnamefont {M.}~\bibnamefont
  {{Eickenberg}}}, \bibinfo {author} {\bibfnamefont {S.}~\bibnamefont {{Ho}}},
  \bibinfo {author} {\bibfnamefont {J.}~\bibnamefont {{Hou}}}, \bibinfo
  {author} {\bibfnamefont {P.}~\bibnamefont {{Lemos}}}, \bibinfo {author}
  {\bibfnamefont {E.}~\bibnamefont {{Massara}}}, \bibinfo {author}
  {\bibfnamefont {A.}~\bibnamefont {{Modi}~Chirag}, \bibfnamefont
  {{Moradinezhad Dizgah}}}, \bibinfo {author} {\bibfnamefont {B.}~\bibnamefont
  {{R\'egaldo-Saint Blancard}}},\ and\ \bibinfo {author} {\bibfnamefont
  {M.~M.}\ \bibnamefont {{Abidi}}},\ }\bibfield  {title} {\bibinfo {title}
  {{{\sc SimBIG}: A Forward Modeling Approach To Analyzing Galaxy
  Clustering}},\ }\href@noop {} {\  (\bibinfo {year} {2022})}\BibitemShut
  {NoStop}%
\bibitem [{\citenamefont {Cranmer}\ \emph {et~al.}(2020)\citenamefont
  {Cranmer}, \citenamefont {Brehmer},\ and\ \citenamefont
  {Louppe}}]{cranmer2020}%
  \BibitemOpen
  \bibfield  {author} {\bibinfo {author} {\bibfnamefont {K.}~\bibnamefont
  {Cranmer}}, \bibinfo {author} {\bibfnamefont {J.}~\bibnamefont {Brehmer}},\
  and\ \bibinfo {author} {\bibfnamefont {G.}~\bibnamefont {Louppe}},\
  }\bibfield  {title} {\bibinfo {title} {The frontier of simulation-based
  inference},\ }\href {https://doi.org/10.1073/pnas.1912789117} {\bibfield
  {journal} {\bibinfo  {journal} {Proceedings of the National Academy of
  Sciences}\ }\textbf {\bibinfo {volume} {117}},\ \bibinfo {pages} {30055}
  (\bibinfo {year} {2020})},\ \Eprint
  {https://arxiv.org/abs/https://www.pnas.org/content/117/48/30055.full.pdf}
  {https://www.pnas.org/content/117/48/30055.full.pdf} \BibitemShut {NoStop}%
\bibitem [{\citenamefont {{Villaescusa-Navarro}}\ \emph
  {et~al.}(2020)\citenamefont {{Villaescusa-Navarro}}, \citenamefont {Hahn},
  \citenamefont {Massara}, \citenamefont {Banerjee}, \citenamefont {Delgado},
  \citenamefont {Ramanah}, \citenamefont {Charnock}, \citenamefont {Giusarma},
  \citenamefont {Li}, \citenamefont {Allys} \emph
  {et~al.}}]{villaescusa-navarro2020}%
  \BibitemOpen
  \bibfield  {author} {\bibinfo {author} {\bibfnamefont {F.}~\bibnamefont
  {{Villaescusa-Navarro}}}, \bibinfo {author} {\bibfnamefont {C.}~\bibnamefont
  {Hahn}}, \bibinfo {author} {\bibfnamefont {E.}~\bibnamefont {Massara}},
  \bibinfo {author} {\bibfnamefont {A.}~\bibnamefont {Banerjee}}, \bibinfo
  {author} {\bibfnamefont {A.~M.}\ \bibnamefont {Delgado}}, \bibinfo {author}
  {\bibfnamefont {D.~K.}\ \bibnamefont {Ramanah}}, \bibinfo {author}
  {\bibfnamefont {T.}~\bibnamefont {Charnock}}, \bibinfo {author}
  {\bibfnamefont {E.}~\bibnamefont {Giusarma}}, \bibinfo {author}
  {\bibfnamefont {Y.}~\bibnamefont {Li}}, \bibinfo {author} {\bibfnamefont
  {E.}~\bibnamefont {Allys}}, \emph {et~al.},\ }\bibfield  {title} {\bibinfo
  {title} {The {{Quijote Simulations}}},\ }\href
  {https://doi.org/10.3847/1538-4365/ab9d82} {\bibfield  {journal} {\bibinfo
  {journal} {The Astrophysical Journal Supplement Series}\ }\textbf {\bibinfo
  {volume} {250}},\ \bibinfo {pages} {2} (\bibinfo {year} {2020})}\BibitemShut
  {NoStop}%
\bibitem [{\citenamefont {Dawson}\ \emph {et~al.}(2013)\citenamefont {Dawson},
  \citenamefont {Schlegel}, \citenamefont {Ahn}, \citenamefont {Anderson},
  \citenamefont {Aubourg}, \citenamefont {Bailey}, \citenamefont {Barkhouser},
  \citenamefont {Bautista}, \citenamefont {Beifiori}, \citenamefont {Berlind}
  \emph {et~al.}}]{dawson2013}%
  \BibitemOpen
  \bibfield  {author} {\bibinfo {author} {\bibfnamefont {K.~S.}\ \bibnamefont
  {Dawson}}, \bibinfo {author} {\bibfnamefont {D.~J.}\ \bibnamefont
  {Schlegel}}, \bibinfo {author} {\bibfnamefont {C.~P.}\ \bibnamefont {Ahn}},
  \bibinfo {author} {\bibfnamefont {S.~F.}\ \bibnamefont {Anderson}}, \bibinfo
  {author} {\bibfnamefont {{\'E}.}~\bibnamefont {Aubourg}}, \bibinfo {author}
  {\bibfnamefont {S.}~\bibnamefont {Bailey}}, \bibinfo {author} {\bibfnamefont
  {R.~H.}\ \bibnamefont {Barkhouser}}, \bibinfo {author} {\bibfnamefont
  {J.~E.}\ \bibnamefont {Bautista}}, \bibinfo {author} {\bibfnamefont
  {A.}~\bibnamefont {Beifiori}}, \bibinfo {author} {\bibfnamefont {A.~A.}\
  \bibnamefont {Berlind}}, \emph {et~al.},\ }\bibfield  {title} {\bibinfo
  {title} {The {{Baryon Oscillation Spectroscopic Survey}} of {{SDSS-III}}},\
  }\href {https://doi.org/10.1088/0004-6256/145/1/10} {\bibfield  {journal}
  {\bibinfo  {journal} {The Astronomical Journal}\ }\textbf {\bibinfo {volume}
  {145}},\ \bibinfo {pages} {10} (\bibinfo {year} {2013})}\BibitemShut
  {NoStop}%
\bibitem [{\citenamefont {Cabass}\ \emph {et~al.}(2023)\citenamefont {Cabass},
  \citenamefont {Simonovi\'c},\ and\ \citenamefont
  {Zaldarriaga}}]{Cabass:2023nyo}%
  \BibitemOpen
  \bibfield  {author} {\bibinfo {author} {\bibfnamefont {G.}~\bibnamefont
  {Cabass}}, \bibinfo {author} {\bibfnamefont {M.}~\bibnamefont
  {Simonovi\'c}},\ and\ \bibinfo {author} {\bibfnamefont {M.}~\bibnamefont
  {Zaldarriaga}},\ }\bibfield  {title} {\bibinfo {title} {{Cosmological
  Information in Perturbative Forward Modeling}},\ }\href@noop {} {\  (\bibinfo
  {year} {2023})},\ \Eprint {https://arxiv.org/abs/2307.04706}
  {arXiv:2307.04706 [astro-ph.CO]} \BibitemShut {NoStop}%
\bibitem [{\citenamefont {Ivanov}\ \emph {et~al.}(2020)\citenamefont {Ivanov},
  \citenamefont {Simonovi{\'c}},\ and\ \citenamefont
  {Zaldarriaga}}]{ivanov2020}%
  \BibitemOpen
  \bibfield  {author} {\bibinfo {author} {\bibfnamefont {M.~M.}\ \bibnamefont
  {Ivanov}}, \bibinfo {author} {\bibfnamefont {M.}~\bibnamefont
  {Simonovi{\'c}}},\ and\ \bibinfo {author} {\bibfnamefont {M.}~\bibnamefont
  {Zaldarriaga}},\ }\bibfield  {title} {\bibinfo {title} {Cosmological
  parameters from the {{BOSS}} galaxy power spectrum},\ }\href
  {https://doi.org/10.1088/1475-7516/2020/05/042} {\bibfield  {journal}
  {\bibinfo  {journal} {Journal of Cosmology and Astroparticle Physics}\
  }\textbf {\bibinfo {volume} {2020}},\ \bibinfo {pages} {042}}\BibitemShut
  {NoStop}%
\bibitem [{\citenamefont {Kokron}\ \emph {et~al.}(2021)\citenamefont {Kokron},
  \citenamefont {DeRose}, \citenamefont {Chen}, \citenamefont {White},\ and\
  \citenamefont {Wechsler}}]{Kokron:2021xgh}%
  \BibitemOpen
  \bibfield  {author} {\bibinfo {author} {\bibfnamefont {N.}~\bibnamefont
  {Kokron}}, \bibinfo {author} {\bibfnamefont {J.}~\bibnamefont {DeRose}},
  \bibinfo {author} {\bibfnamefont {S.-F.}\ \bibnamefont {Chen}}, \bibinfo
  {author} {\bibfnamefont {M.}~\bibnamefont {White}},\ and\ \bibinfo {author}
  {\bibfnamefont {R.~H.}\ \bibnamefont {Wechsler}},\ }\bibfield  {title}
  {\bibinfo {title} {{The cosmology dependence of galaxy clustering and lensing
  from a hybrid N-body\textendash{}perturbation theory model}},\ }\href
  {https://doi.org/10.1093/mnras/stab1358} {\bibfield  {journal} {\bibinfo
  {journal} {Mon. Not. Roy. Astron. Soc.}\ }\textbf {\bibinfo {volume} {505}},\
  \bibinfo {pages} {1422} (\bibinfo {year} {2021})},\ \Eprint
  {https://arxiv.org/abs/2101.11014} {arXiv:2101.11014 [astro-ph.CO]}
  \BibitemShut {NoStop}%
\bibitem [{\citenamefont {{Modi}}\ \emph {et~al.}(2020)\citenamefont {{Modi}},
  \citenamefont {{Chen}},\ and\ \citenamefont {{White}}}]{Modi2020}%
  \BibitemOpen
  \bibfield  {author} {\bibinfo {author} {\bibfnamefont {C.}~\bibnamefont
  {{Modi}}}, \bibinfo {author} {\bibfnamefont {S.-F.}\ \bibnamefont {{Chen}}},\
  and\ \bibinfo {author} {\bibfnamefont {M.}~\bibnamefont {{White}}},\
  }\bibfield  {title} {\bibinfo {title} {{Simulations and symmetries}},\ }\href
  {https://doi.org/10.1093/mnras/staa251} {\bibfield  {journal} {\bibinfo
  {journal} {\mnras}\ }\textbf {\bibinfo {volume} {492}},\ \bibinfo {pages}
  {5754} (\bibinfo {year} {2020})},\ \Eprint {https://arxiv.org/abs/1910.07097}
  {arXiv:1910.07097 [astro-ph.CO]} \BibitemShut {NoStop}%
\bibitem [{\citenamefont {Springel}(2005)}]{Springel:2005mi}%
  \BibitemOpen
  \bibfield  {author} {\bibinfo {author} {\bibfnamefont {V.}~\bibnamefont
  {Springel}},\ }\bibfield  {title} {\bibinfo {title} {{The Cosmological
  simulation code GADGET-2}},\ }\href
  {https://doi.org/10.1111/j.1365-2966.2005.09655.x} {\bibfield  {journal}
  {\bibinfo  {journal} {Mon. Not. Roy. Astron. Soc.}\ }\textbf {\bibinfo
  {volume} {364}},\ \bibinfo {pages} {1105} (\bibinfo {year} {2005})},\ \Eprint
  {https://arxiv.org/abs/astro-ph/0505010} {arXiv:astro-ph/0505010}
  \BibitemShut {NoStop}%
\bibitem [{\citenamefont {Hand}\ \emph {et~al.}(2018)\citenamefont {Hand},
  \citenamefont {Feng}, \citenamefont {Beutler}, \citenamefont {Li},
  \citenamefont {Modi}, \citenamefont {Seljak},\ and\ \citenamefont
  {Slepian}}]{hand2018}%
  \BibitemOpen
  \bibfield  {author} {\bibinfo {author} {\bibfnamefont {N.}~\bibnamefont
  {Hand}}, \bibinfo {author} {\bibfnamefont {Y.}~\bibnamefont {Feng}}, \bibinfo
  {author} {\bibfnamefont {F.}~\bibnamefont {Beutler}}, \bibinfo {author}
  {\bibfnamefont {Y.}~\bibnamefont {Li}}, \bibinfo {author} {\bibfnamefont
  {C.}~\bibnamefont {Modi}}, \bibinfo {author} {\bibfnamefont {U.}~\bibnamefont
  {Seljak}},\ and\ \bibinfo {author} {\bibfnamefont {Z.}~\bibnamefont
  {Slepian}},\ }\bibfield  {title} {\bibinfo {title} {Nbodykit: {{An
  Open-source}}, {{Massively Parallel Toolkit}} for {{Large-scale
  Structure}}},\ }\href {https://doi.org/10.3847/1538-3881/aadae0} {\bibfield
  {journal} {\bibinfo  {journal} {The Astronomical Journal}\ }\textbf {\bibinfo
  {volume} {156}},\ \bibinfo {pages} {160} (\bibinfo {year}
  {2018})}\BibitemShut {NoStop}%
\bibitem [{\citenamefont {Baumann}\ \emph {et~al.}(2012)\citenamefont
  {Baumann}, \citenamefont {Nicolis}, \citenamefont {Senatore},\ and\
  \citenamefont {Zaldarriaga}}]{Baumann:2010tm}%
  \BibitemOpen
  \bibfield  {author} {\bibinfo {author} {\bibfnamefont {D.}~\bibnamefont
  {Baumann}}, \bibinfo {author} {\bibfnamefont {A.}~\bibnamefont {Nicolis}},
  \bibinfo {author} {\bibfnamefont {L.}~\bibnamefont {Senatore}},\ and\
  \bibinfo {author} {\bibfnamefont {M.}~\bibnamefont {Zaldarriaga}},\
  }\bibfield  {title} {\bibinfo {title} {{Cosmological Non-Linearities as an
  Effective Fluid}},\ }\href {https://doi.org/10.1088/1475-7516/2012/07/051}
  {\bibfield  {journal} {\bibinfo  {journal} {JCAP}\ }\textbf {\bibinfo
  {volume} {07}},\ \bibinfo {pages} {051}},\ \Eprint
  {https://arxiv.org/abs/1004.2488} {arXiv:1004.2488 [astro-ph.CO]}
  \BibitemShut {NoStop}%
\bibitem [{\citenamefont {Carrasco}\ \emph {et~al.}(2012)\citenamefont
  {Carrasco}, \citenamefont {Hertzberg},\ and\ \citenamefont
  {Senatore}}]{Carrasco:2012cv}%
  \BibitemOpen
  \bibfield  {author} {\bibinfo {author} {\bibfnamefont {J.~J.~M.}\
  \bibnamefont {Carrasco}}, \bibinfo {author} {\bibfnamefont {M.~P.}\
  \bibnamefont {Hertzberg}},\ and\ \bibinfo {author} {\bibfnamefont
  {L.}~\bibnamefont {Senatore}},\ }\bibfield  {title} {\bibinfo {title} {{The
  Effective Field Theory of Cosmological Large Scale Structures}},\ }\href
  {https://doi.org/10.1007/JHEP09(2012)082} {\bibfield  {journal} {\bibinfo
  {journal} {JHEP}\ }\textbf {\bibinfo {volume} {09}},\ \bibinfo {pages}
  {082}},\ \Eprint {https://arxiv.org/abs/1206.2926} {arXiv:1206.2926
  [astro-ph.CO]} \BibitemShut {NoStop}%
\bibitem [{\citenamefont {Chudaykin}\ \emph {et~al.}(2020)\citenamefont
  {Chudaykin}, \citenamefont {Ivanov}, \citenamefont {Philcox},\ and\
  \citenamefont {Simonovi\'c}}]{Chudaykin:2020aoj}%
  \BibitemOpen
  \bibfield  {author} {\bibinfo {author} {\bibfnamefont {A.}~\bibnamefont
  {Chudaykin}}, \bibinfo {author} {\bibfnamefont {M.~M.}\ \bibnamefont
  {Ivanov}}, \bibinfo {author} {\bibfnamefont {O.~H.~E.}\ \bibnamefont
  {Philcox}},\ and\ \bibinfo {author} {\bibfnamefont {M.}~\bibnamefont
  {Simonovi\'c}},\ }\bibfield  {title} {\bibinfo {title} {{Nonlinear
  perturbation theory extension of the Boltzmann code CLASS}},\ }\href
  {https://doi.org/10.1103/PhysRevD.102.063533} {\bibfield  {journal} {\bibinfo
   {journal} {Phys. Rev. D}\ }\textbf {\bibinfo {volume} {102}},\ \bibinfo
  {pages} {063533} (\bibinfo {year} {2020})},\ \Eprint
  {https://arxiv.org/abs/2004.10607} {arXiv:2004.10607 [astro-ph.CO]}
  \BibitemShut {NoStop}%
\bibitem [{\citenamefont {{Lanusse}}\ \emph {et~al.}(2012)\citenamefont
  {{Lanusse}}, \citenamefont {{Rassat}},\ and\ \citenamefont
  {{Starck}}}]{Lanusse2012}%
  \BibitemOpen
  \bibfield  {author} {\bibinfo {author} {\bibfnamefont {F.}~\bibnamefont
  {{Lanusse}}}, \bibinfo {author} {\bibfnamefont {A.}~\bibnamefont
  {{Rassat}}},\ and\ \bibinfo {author} {\bibfnamefont {J.~L.}\ \bibnamefont
  {{Starck}}},\ }\bibfield  {title} {\bibinfo {title} {{Spherical 3D isotropic
  wavelets}},\ }\href {https://doi.org/10.1051/0004-6361/201118568} {\bibfield
  {journal} {\bibinfo  {journal} {\aap}\ }\textbf {\bibinfo {volume} {540}},\
  \bibinfo {eid} {A92} (\bibinfo {year} {2012})},\ \Eprint
  {https://arxiv.org/abs/1112.0561} {arXiv:1112.0561 [astro-ph.CO]}
  \BibitemShut {NoStop}%
\bibitem [{\citenamefont {{Eickenberg}}\ \emph {et~al.}(2022)\citenamefont
  {{Eickenberg}}, \citenamefont {{Allys}}, \citenamefont {{Moradinezhad
  Dizgah}}, \citenamefont {{Lemos}}, \citenamefont {{Massara}}, \citenamefont
  {{Abidi}}, \citenamefont {{Hahn}}, \citenamefont {{Hassan}}, \citenamefont
  {{Regaldo-Saint Blancard}}, \citenamefont {{Ho}}, \citenamefont {{Mallat}},
  \citenamefont {{And{\'e}n}},\ and\ \citenamefont
  {{Villaescusa-Navarro}}}]{Eickenberg2022}%
  \BibitemOpen
  \bibfield  {author} {\bibinfo {author} {\bibfnamefont {M.}~\bibnamefont
  {{Eickenberg}}}, \bibinfo {author} {\bibfnamefont {E.}~\bibnamefont
  {{Allys}}}, \bibinfo {author} {\bibfnamefont {A.}~\bibnamefont {{Moradinezhad
  Dizgah}}}, \bibinfo {author} {\bibfnamefont {P.}~\bibnamefont {{Lemos}}},
  \bibinfo {author} {\bibfnamefont {E.}~\bibnamefont {{Massara}}}, \bibinfo
  {author} {\bibfnamefont {M.}~\bibnamefont {{Abidi}}}, \bibinfo {author}
  {\bibfnamefont {C.}~\bibnamefont {{Hahn}}}, \bibinfo {author} {\bibfnamefont
  {S.}~\bibnamefont {{Hassan}}}, \bibinfo {author} {\bibfnamefont
  {B.}~\bibnamefont {{Regaldo-Saint Blancard}}}, \bibinfo {author}
  {\bibfnamefont {S.}~\bibnamefont {{Ho}}}, \bibinfo {author} {\bibfnamefont
  {S.}~\bibnamefont {{Mallat}}}, \bibinfo {author} {\bibfnamefont
  {J.}~\bibnamefont {{And{\'e}n}}},\ and\ \bibinfo {author} {\bibfnamefont
  {F.}~\bibnamefont {{Villaescusa-Navarro}}},\ }\bibfield  {title} {\bibinfo
  {title} {{Wavelet Moments for Cosmological Parameter Estimation}},\ }\href
  {https://doi.org/10.48550/arXiv.2204.07646} {\bibfield  {journal} {\bibinfo
  {journal} {arXiv e-prints}\ ,\ \bibinfo {eid} {arXiv:2204.07646}} (\bibinfo
  {year} {2022})},\ \Eprint {https://arxiv.org/abs/2204.07646}
  {arXiv:2204.07646 [astro-ph.CO]} \BibitemShut {NoStop}%
\bibitem [{\citenamefont {Foreman-Mackey}\ \emph {et~al.}(2013)\citenamefont
  {Foreman-Mackey}, \citenamefont {Hogg}, \citenamefont {Lang},\ and\
  \citenamefont {Goodman}}]{Foreman-Mackey:2012any}%
  \BibitemOpen
  \bibfield  {author} {\bibinfo {author} {\bibfnamefont {D.}~\bibnamefont
  {Foreman-Mackey}}, \bibinfo {author} {\bibfnamefont {D.~W.}\ \bibnamefont
  {Hogg}}, \bibinfo {author} {\bibfnamefont {D.}~\bibnamefont {Lang}},\ and\
  \bibinfo {author} {\bibfnamefont {J.}~\bibnamefont {Goodman}},\ }\bibfield
  {title} {\bibinfo {title} {{emcee: The MCMC Hammer}},\ }\href
  {https://doi.org/10.1086/670067} {\bibfield  {journal} {\bibinfo  {journal}
  {Publ. Astron. Soc. Pac.}\ }\textbf {\bibinfo {volume} {125}},\ \bibinfo
  {pages} {306} (\bibinfo {year} {2013})},\ \Eprint
  {https://arxiv.org/abs/1202.3665} {arXiv:1202.3665 [astro-ph.IM]}
  \BibitemShut {NoStop}%
\bibitem [{\citenamefont {Tassev}\ \emph {et~al.}(2015)\citenamefont {Tassev},
  \citenamefont {Eisenstein}, \citenamefont {Wandelt},\ and\ \citenamefont
  {Zaldarriaga}}]{Tassev:2015mia}%
  \BibitemOpen
  \bibfield  {author} {\bibinfo {author} {\bibfnamefont {S.}~\bibnamefont
  {Tassev}}, \bibinfo {author} {\bibfnamefont {D.~J.}\ \bibnamefont
  {Eisenstein}}, \bibinfo {author} {\bibfnamefont {B.~D.}\ \bibnamefont
  {Wandelt}},\ and\ \bibinfo {author} {\bibfnamefont {M.}~\bibnamefont
  {Zaldarriaga}},\ }\bibfield  {title} {\bibinfo {title} {{sCOLA: The N-body
  COLA Method Extended to the Spatial Domain}},\ }\href@noop {} {\  (\bibinfo
  {year} {2015})},\ \Eprint {https://arxiv.org/abs/1502.07751}
  {arXiv:1502.07751 [astro-ph.CO]} \BibitemShut {NoStop}%
\bibitem [{\citenamefont {Leclercq}\ \emph {et~al.}(2020)\citenamefont
  {Leclercq}, \citenamefont {Faure}, \citenamefont {Lavaux}, \citenamefont
  {Wandelt}, \citenamefont {Jaffe}, \citenamefont {Heavens}, \citenamefont
  {Percival},\ and\ \citenamefont {No\^us}}]{Leclercq:2020nmz}%
  \BibitemOpen
  \bibfield  {author} {\bibinfo {author} {\bibfnamefont {F.}~\bibnamefont
  {Leclercq}}, \bibinfo {author} {\bibfnamefont {B.}~\bibnamefont {Faure}},
  \bibinfo {author} {\bibfnamefont {G.}~\bibnamefont {Lavaux}}, \bibinfo
  {author} {\bibfnamefont {B.~D.}\ \bibnamefont {Wandelt}}, \bibinfo {author}
  {\bibfnamefont {A.~H.}\ \bibnamefont {Jaffe}}, \bibinfo {author}
  {\bibfnamefont {A.~F.}\ \bibnamefont {Heavens}}, \bibinfo {author}
  {\bibfnamefont {W.~J.}\ \bibnamefont {Percival}},\ and\ \bibinfo {author}
  {\bibfnamefont {C.}~\bibnamefont {No\^us}},\ }\bibfield  {title} {\bibinfo
  {title} {{Perfectly parallel cosmological simulations using spatial comoving
  Lagrangian acceleration}},\ }\href
  {https://doi.org/10.1051/0004-6361/202037995} {\bibfield  {journal} {\bibinfo
   {journal} {Astron. Astrophys.}\ }\textbf {\bibinfo {volume} {639}},\
  \bibinfo {pages} {A91} (\bibinfo {year} {2020})},\ \Eprint
  {https://arxiv.org/abs/2003.04925} {arXiv:2003.04925 [astro-ph.CO]}
  \BibitemShut {NoStop}%
\bibitem [{\citenamefont {Modi}\ \emph {et~al.}(2021)\citenamefont {Modi},
  \citenamefont {Lanusse},\ and\ \citenamefont {Seljak}}]{modi2021}%
  \BibitemOpen
  \bibfield  {author} {\bibinfo {author} {\bibfnamefont {C.}~\bibnamefont
  {Modi}}, \bibinfo {author} {\bibfnamefont {F.}~\bibnamefont {Lanusse}},\ and\
  \bibinfo {author} {\bibfnamefont {U.}~\bibnamefont {Seljak}},\ }\bibfield
  {title} {\bibinfo {title} {{{FlowPM}}: {{Distributed TensorFlow}}
  implementation of the {{FastPM}} cosmological {{N-body}} solver},\ }\href
  {https://doi.org/10.1016/j.ascom.2021.100505} {\bibfield  {journal} {\bibinfo
   {journal} {Astronomy and Computing, Volume 37, article id. 100505.}\
  }\textbf {\bibinfo {volume} {37}},\ \bibinfo {pages} {100505} (\bibinfo
  {year} {2021})}\BibitemShut {NoStop}%
\bibitem [{\citenamefont {{Nadler}}\ \emph {et~al.}(2023)\citenamefont
  {{Nadler}}, \citenamefont {{Mansfield}}, \citenamefont {{Wang}},
  \citenamefont {{Du}}, \citenamefont {{Adhikari}}, \citenamefont {{Banerjee}},
  \citenamefont {{Benson}}, \citenamefont {{Darragh-Ford}}, \citenamefont
  {{Mao}}, \citenamefont {{Wagner-Carena}}, \citenamefont {{Wechsler}},\ and\
  \citenamefont {{Wu}}}]{Nadler2023}%
  \BibitemOpen
  \bibfield  {author} {\bibinfo {author} {\bibfnamefont {E.~O.}\ \bibnamefont
  {{Nadler}}}, \bibinfo {author} {\bibfnamefont {P.}~\bibnamefont
  {{Mansfield}}}, \bibinfo {author} {\bibfnamefont {Y.}~\bibnamefont {{Wang}}},
  \bibinfo {author} {\bibfnamefont {X.}~\bibnamefont {{Du}}}, \bibinfo {author}
  {\bibfnamefont {S.}~\bibnamefont {{Adhikari}}}, \bibinfo {author}
  {\bibfnamefont {A.}~\bibnamefont {{Banerjee}}}, \bibinfo {author}
  {\bibfnamefont {A.}~\bibnamefont {{Benson}}}, \bibinfo {author}
  {\bibfnamefont {E.}~\bibnamefont {{Darragh-Ford}}}, \bibinfo {author}
  {\bibfnamefont {Y.-Y.}\ \bibnamefont {{Mao}}}, \bibinfo {author}
  {\bibfnamefont {S.}~\bibnamefont {{Wagner-Carena}}}, \bibinfo {author}
  {\bibfnamefont {R.~H.}\ \bibnamefont {{Wechsler}}},\ and\ \bibinfo {author}
  {\bibfnamefont {H.-Y.}\ \bibnamefont {{Wu}}},\ }\bibfield  {title} {\bibinfo
  {title} {{Symphony: Cosmological Zoom-in Simulation Suites over Four Decades
  of Host Halo Mass}},\ }\href {https://doi.org/10.3847/1538-4357/acb68c}
  {\bibfield  {journal} {\bibinfo  {journal} {\apj}\ }\textbf {\bibinfo
  {volume} {945}},\ \bibinfo {eid} {159} (\bibinfo {year} {2023})},\ \Eprint
  {https://arxiv.org/abs/2209.02675} {arXiv:2209.02675 [astro-ph.CO]}
  \BibitemShut {NoStop}%
\bibitem [{\citenamefont {Li}\ \emph {et~al.}(2022)\citenamefont {Li},
  \citenamefont {Lu}, \citenamefont {Modi}, \citenamefont {Jamieson},
  \citenamefont {Zhang}, \citenamefont {Feng}, \citenamefont {Zhou},
  \citenamefont {Kwan}, \citenamefont {Lanusse},\ and\ \citenamefont
  {Greengard}}]{Li:2022qlf}%
  \BibitemOpen
  \bibfield  {author} {\bibinfo {author} {\bibfnamefont {Y.}~\bibnamefont
  {Li}}, \bibinfo {author} {\bibfnamefont {L.}~\bibnamefont {Lu}}, \bibinfo
  {author} {\bibfnamefont {C.}~\bibnamefont {Modi}}, \bibinfo {author}
  {\bibfnamefont {D.}~\bibnamefont {Jamieson}}, \bibinfo {author}
  {\bibfnamefont {Y.}~\bibnamefont {Zhang}}, \bibinfo {author} {\bibfnamefont
  {Y.}~\bibnamefont {Feng}}, \bibinfo {author} {\bibfnamefont {W.}~\bibnamefont
  {Zhou}}, \bibinfo {author} {\bibfnamefont {N.~P.}\ \bibnamefont {Kwan}},
  \bibinfo {author} {\bibfnamefont {F.}~\bibnamefont {Lanusse}},\ and\ \bibinfo
  {author} {\bibfnamefont {L.}~\bibnamefont {Greengard}},\ }\bibfield  {title}
  {\bibinfo {title} {{pmwd: A Differentiable Cosmological Particle-Mesh
  $N$-body Library}},\ }\href@noop {} {\  (\bibinfo {year} {2022})},\ \Eprint
  {https://arxiv.org/abs/2211.09958} {arXiv:2211.09958 [astro-ph.IM]}
  \BibitemShut {NoStop}%
\bibitem [{\citenamefont {Goldstein}\ \emph {et~al.}(2022)\citenamefont
  {Goldstein}, \citenamefont {Esposito}, \citenamefont {Philcox}, \citenamefont
  {Hui}, \citenamefont {Hill}, \citenamefont {Scoccimarro},\ and\ \citenamefont
  {Abitbol}}]{Goldstein:2022hgr}%
  \BibitemOpen
  \bibfield  {author} {\bibinfo {author} {\bibfnamefont {S.}~\bibnamefont
  {Goldstein}}, \bibinfo {author} {\bibfnamefont {A.}~\bibnamefont {Esposito}},
  \bibinfo {author} {\bibfnamefont {O.~H.~E.}\ \bibnamefont {Philcox}},
  \bibinfo {author} {\bibfnamefont {L.}~\bibnamefont {Hui}}, \bibinfo {author}
  {\bibfnamefont {J.~C.}\ \bibnamefont {Hill}}, \bibinfo {author}
  {\bibfnamefont {R.}~\bibnamefont {Scoccimarro}},\ and\ \bibinfo {author}
  {\bibfnamefont {M.~H.}\ \bibnamefont {Abitbol}},\ }\bibfield  {title}
  {\bibinfo {title} {{Squeezing fNL out of the matter bispectrum with
  consistency relations}},\ }\href
  {https://doi.org/10.1103/PhysRevD.106.123525} {\bibfield  {journal} {\bibinfo
   {journal} {Phys. Rev. D}\ }\textbf {\bibinfo {volume} {106}},\ \bibinfo
  {pages} {123525} (\bibinfo {year} {2022})},\ \Eprint
  {https://arxiv.org/abs/2209.06228} {arXiv:2209.06228 [astro-ph.CO]}
  \BibitemShut {NoStop}%
\bibitem [{\citenamefont {Cabass}\ \emph
  {et~al.}(2022{\natexlab{a}})\citenamefont {Cabass}, \citenamefont {Ivanov},
  \citenamefont {Philcox}, \citenamefont {Simonovi\'c},\ and\ \citenamefont
  {Zaldarriaga}}]{Cabass:2022wjy}%
  \BibitemOpen
  \bibfield  {author} {\bibinfo {author} {\bibfnamefont {G.}~\bibnamefont
  {Cabass}}, \bibinfo {author} {\bibfnamefont {M.~M.}\ \bibnamefont {Ivanov}},
  \bibinfo {author} {\bibfnamefont {O.~H.~E.}\ \bibnamefont {Philcox}},
  \bibinfo {author} {\bibfnamefont {M.}~\bibnamefont {Simonovi\'c}},\ and\
  \bibinfo {author} {\bibfnamefont {M.}~\bibnamefont {Zaldarriaga}},\
  }\bibfield  {title} {\bibinfo {title} {{Constraints on Single-Field Inflation
  from the BOSS Galaxy Survey}},\ }\href
  {https://doi.org/10.1103/PhysRevLett.129.021301} {\bibfield  {journal}
  {\bibinfo  {journal} {Phys. Rev. Lett.}\ }\textbf {\bibinfo {volume} {129}},\
  \bibinfo {pages} {021301} (\bibinfo {year} {2022}{\natexlab{a}})},\ \Eprint
  {https://arxiv.org/abs/2201.07238} {arXiv:2201.07238 [astro-ph.CO]}
  \BibitemShut {NoStop}%
\bibitem [{\citenamefont {Cabass}\ \emph
  {et~al.}(2022{\natexlab{b}})\citenamefont {Cabass}, \citenamefont {Ivanov},
  \citenamefont {Philcox}, \citenamefont {Simonovi\'c},\ and\ \citenamefont
  {Zaldarriaga}}]{Cabass:2022ymb}%
  \BibitemOpen
  \bibfield  {author} {\bibinfo {author} {\bibfnamefont {G.}~\bibnamefont
  {Cabass}}, \bibinfo {author} {\bibfnamefont {M.~M.}\ \bibnamefont {Ivanov}},
  \bibinfo {author} {\bibfnamefont {O.~H.~E.}\ \bibnamefont {Philcox}},
  \bibinfo {author} {\bibfnamefont {M.}~\bibnamefont {Simonovi\'c}},\ and\
  \bibinfo {author} {\bibfnamefont {M.}~\bibnamefont {Zaldarriaga}},\
  }\bibfield  {title} {\bibinfo {title} {{Constraints on multifield inflation
  from the BOSS galaxy survey}},\ }\href
  {https://doi.org/10.1103/PhysRevD.106.043506} {\bibfield  {journal} {\bibinfo
   {journal} {Phys. Rev. D}\ }\textbf {\bibinfo {volume} {106}},\ \bibinfo
  {pages} {043506} (\bibinfo {year} {2022}{\natexlab{b}})},\ \Eprint
  {https://arxiv.org/abs/2204.01781} {arXiv:2204.01781 [astro-ph.CO]}
  \BibitemShut {NoStop}%
\bibitem [{\citenamefont {D'Amico}\ \emph
  {et~al.}(2022{\natexlab{b}})\citenamefont {D'Amico}, \citenamefont
  {Lewandowski}, \citenamefont {Senatore},\ and\ \citenamefont
  {Zhang}}]{DAmico:2022gki}%
  \BibitemOpen
  \bibfield  {author} {\bibinfo {author} {\bibfnamefont {G.}~\bibnamefont
  {D'Amico}}, \bibinfo {author} {\bibfnamefont {M.}~\bibnamefont
  {Lewandowski}}, \bibinfo {author} {\bibfnamefont {L.}~\bibnamefont
  {Senatore}},\ and\ \bibinfo {author} {\bibfnamefont {P.}~\bibnamefont
  {Zhang}},\ }\bibfield  {title} {\bibinfo {title} {{Limits on primordial
  non-Gaussianities from BOSS galaxy-clustering data}},\ }\href@noop {} {\
  (\bibinfo {year} {2022}{\natexlab{b}})},\ \Eprint
  {https://arxiv.org/abs/2201.11518} {arXiv:2201.11518 [astro-ph.CO]}
  \BibitemShut {NoStop}%
\end{thebibliography}%

\end{document}